%% file: GCMB.tex
\documentclass[journal,12pt,onecolumn]{IEEEtran}

\usepackage{setspace}
\ifCLASSOPTIONonecolumn \doublespace \fi
\usepackage[latin1]{inputenc}
\usepackage{amsmath}
\usepackage{amsfonts}
\usepackage{amssymb}
\usepackage{graphicx}
\usepackage{cite}
\newtheorem{theorem}{Theorem}

\ifCLASSOPTIONonecolumn
\title{Golden Coded Multiple Beamforming}
\author{
Boyu Li and Ender Ayanoglu\\
Center for Pervasive Communications and Computing\\ 
Department of Electrical Engineering and Computer Science\\
The Henry Samueli School of Engineering\\
University of California, Irvine\\
Irvine, California 92697-2625\\
Email: boyul@uci.edu, ayanoglu@uci.edu}
\date{} 
\else
\title{Golden Coded Multiple Beamforming}
\author{
Boyu Li and Ender Ayanoglu\\
\\
Center for Pervasive Communications and Computing\\ 
Department of Electrical Engineering and Computer Science\\
The Henry Samueli School of Engineering\\
University of California, Irvine\\
Irvine, California 92697-2625\\
Email: boyul@uci.edu, ayanoglu@uci.edu}
\date{} 
\fi

\begin{document}
\maketitle

\ifCLASSOPTIONonecolumn
 \setlength\arraycolsep{4pt}
\else
 \setlength\arraycolsep{2pt}
\fi

\input{Abstract}
\ifCLASSOPTIONonecolumn \newpage \fi
\input{Introduction}
\input{System_model}
\input{Analysis}
\input{Decoding}
\input{Extension}
\input{Results}
\input{Conclusion}
\ifCLASSOPTIONonecolumn \newpage \fi
\bibliographystyle{IEEEtran}
\bibliography{IEEEabrv,Mybib}
\end{document}

%% file: Abstract.tex
\begin{abstract}

The Golden Code is a full-rate full-diversity space-time code, which achieves maximum coding gain for Multiple-Input Multiple-Output (MIMO) systems with two transmit and two receive antennas. Since four information symbols taken from an $M$-QAM constellation are selected to construct one Golden Code codeword, a maximum likelihood decoder using sphere decoding has the worst-case complexity of $\mathcal{O}(M^4)$, when the Channel State Information (CSI) is available at the receiver. Previously, this worst-case complexity was reduced to $\mathcal{O}(M^{2.5})$ without performance degradation. When the CSI is known by the transmitter as well as the receiver, beamforming techniques that employ singular value decomposition are commonly used in MIMO systems. In the absence of channel coding, when a single symbol is transmitted, these systems achieve the full diversity order provided by the channel. Whereas this property is lost when multiple symbols are simultaneously transmitted. However, uncoded multiple beamforming can achieve the full diversity order by adding a properly designed constellation precoder. For $2\times2$ Fully Precoded Multiple Beamforming (FPMB), the general worst-case decoding complexity is $\mathcal{O}(M)$. In this paper, Golden Coded Multiple Beamforming (GCMB) is proposed, which transmits the Golden Code through $2\times2$ multiple beamforming. GCMB achieves the full diversity order and its performance is similar to general MIMO systems using the Golden Code and FPMB, whereas the worst-case decoding complexity of $\mathcal{O}(\sqrt{M})$ is much lower. The extension of GCMB to larger dimensions is also discussed.

\end{abstract}

%% file: Introduction.tex
\section{Introduction} \label{sec:Introduction}


The Golden Code is a space-time code for Multiple-Input Multiple-Output (MIMO) systems with two transmit and two receive antennas \cite{Belfiore_GC}, \cite{Dayal_STC}. It achieves both the full rate and the full diversity. Its Bit Error Rate (BER) performance is so far the best compared to other $2\times2$ space-time codes. Moreover, it has a nonvanishing coding gain that is independent of the size of the signal constellation, thus it achieves the optimal diversity-multiplexing performance presented in \cite{Zheng_DM}. Because of those advantages, the Golden Code has been incorporated into the $802.16$e WiMAX standard \cite{IEEE_802_16e}.

Since each Golden Code codeword employs four information symbols from an $M$-QAM constellation, $M^4$ points are calculated by exhaustive search to achieve the Maximum Likelihood (ML) decoding. Hence, the decoding complexity is proportional to $M^4$, denoted by $\mathcal{O}(M^4)$. Sphere Decoding (SD) is an alternative for ML with reduced complexity \cite{Jalden_SD}. While SD reduces the average decoding complexity, the worst-case complexity is still $\mathcal{O}(M^4)$.

To reduce the decoding complexity of the Golden Code, several techniques have been proposed. In \cite{Sinnokrot_FMLD_GC}, \cite{Sinnokrot_GC_FD}, the worst-case complexity of the Golden Code is reduced to $\mathcal{O}(M^{2.5})$ without performance degradation. In \cite{Zhang_GC_SD}, an improved sphere decoding for the Golden Code is designed to reduce the average decoding complexity. In \cite{Howard_FD_GC}, a decoding technique with the complexity of $\mathcal{O}(M^{2})$ is presented, which is based on the Diophantine approximation and with the trade-off of $2$dB performance loss. Other suboptimal decoders for the Golden Code are discussed in \cite{Sarkiss_PC_GC} and \cite{Othman_GC}.

When channel state information (CSI) is available at the transmitter, beamforming techniques, which exploit Singular Value Decomposition (SVD), are applied in a MIMO system to achieve spatial multiplexing and thereby increase the data rate, or to enhance the performance \cite{Jafarkhani_STC}. However, spatial multiplexing without channel coding results in the loss of the full diversity order \cite{Sengul_DA_SMB}. To overcome the diversity degradation of multiple beamforming, the constellation precoding technique can be employed \cite{Park_CPB}. It is shown in \cite{Park_CPB} that Fully Precoded Multiple Beamforming (FPMB) achieves full diversity.  

In this paper, the technique of Golden Coded Multiple Beamforming (GCMB) that combines the Golden Code with $2\times2$ multiple beamforming is proposed. GCMB achieves both the full rate and the full diversity similar to the general MIMO systems employing the Golden Code and $2\times2$ FPMB. All these three techniques have almost the same BER performance. However, the worst-case decoding complexity of GCMB is reduced to $\mathcal{O}(\sqrt{M})$ compared to general MIMO systems using the Golden Code. This complexity is lower than FPMB as well, whose worst-case complexity is $\mathcal{O}(M)$. 

The remainder of this paper is organized as follows: In Section \ref{sec:System_model}, the description of GCMB is given. In Section \ref{sec:Analysis}, the diversity analysis of GCMB is provided. In Section \ref{sec:Decoding}, the decoding technique and complexity of GCMB are shown. In Section \ref{sec:Extension}, the extension of GCMB to larger dimensions is discussed. In Section \ref{sec:Results}, performance comparisons of different techniques are carried out. Finally, a conclusion is provided in Section \ref{sec:Conclusion}.

%% file: System_model.tex
\section{GCMB Overview} \label{sec:System_model}

Fig. \ref{fig:system_model} represents the structure of GCMB. Firstly, the information bit sequence is modulated by the $M$-ary square QAM. Then four consecutive modulated complex-valued scalar symbols $s_1$, $s_2$, $s_3$, and $s_4$ are encoded into the Golden Code codewords. The codewords $\mathbf{X}$ of the Golden Code are $2\times2$ complex-valued matrices \cite{Belfiore_GC}, given as
\begin{align}
\mathbf{X} = \frac{1}{\sqrt{5}} 
\left[ \begin{array}{cc} 
(1+i\beta)s_1+({\alpha}-i)s_2 & (1+i\beta)s_3+({\alpha}-i)s_4 \\
(i-{\alpha})s_3+(1+i\beta)s_4 & (1+i\alpha)s_1+({\beta}-i)s_2
\end{array} \right], 
\label{eq:golden_code} 
\end{align}
with $\alpha=\frac{1+\sqrt{5}}{2}$ and $\beta=(\frac{1-\sqrt{5}}{2})$.  

\ifCLASSOPTIONonecolumn
\begin{figure}[!m]
\centering \includegraphics[width = 0.6\linewidth]{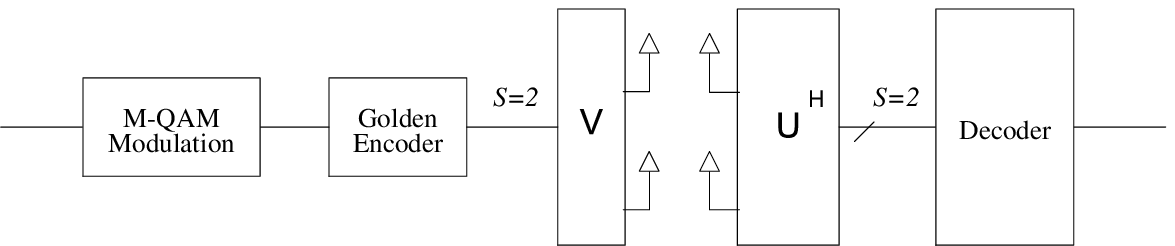}
\caption{Structure of GCMB.} \label{fig:system_model}
\end{figure}
\else
\begin{figure}[!t]
\centering \includegraphics[width = 1.0\linewidth]{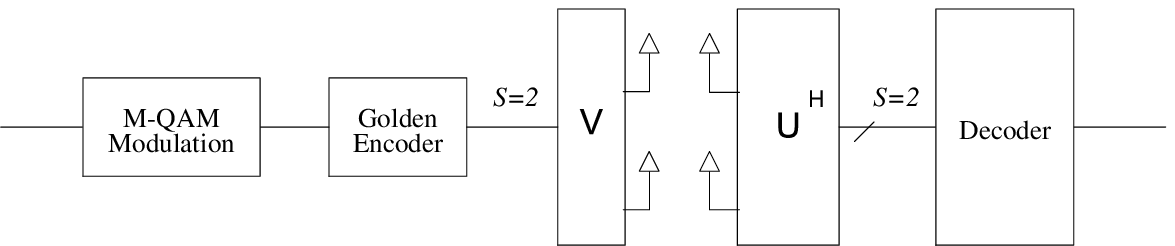}
\caption{Structure of GCMB.} \label{fig:system_model}
\end{figure}
\fi

The MIMO channel $\mathbf{H} \in \mathbb{C}^{N_r \times N_t}$ is assumed to be quasi-static, Rayleigh, and flat fading, and known by both the transmitter and the receiver, where $N_r=N_t=2$ denote the number of transmit and receive antennas respectively, and $\mathbb{C}$ stands for the set of complex numbers. The beamforming vectors are determined by the SVD of the MIMO channel, i.e., $\mathbf{H} = \mathbf{U \Lambda V}^H$ where $\mathbf{U}$ and $\mathbf{V}$ are unitary matrices, and $\mathbf{\Lambda}$ is a diagonal matrix whose $s^{th}$ diagonal element, $\lambda_s \in \mathbb{R}^+$, is a singular value of $\mathbf{H}$ in decreasing order, where $\mathbb{R}^+$ denotes the set of positive real numbers. When $S$ streams are transmitted at the same time, the first $S$ vectors of $\mathbf{U}$ and $\mathbf{V}$ are chosen to be used as beamforming matrices at the receiver and the transmitter, respectively. In the case of GCMB, the number of streams $S=2$.

The received signal is
\begin{align}
\mathbf{Y} = \mathbf{{\Lambda}X}+\mathbf{N}, \label{eq:detected_matrix}
\end{align}
where $\mathbf{Y}$ is a $2\times2$ complex-valued matrix, and $\mathbf{N}$ is the $2\times2$ complex-valued additive white Gaussian noise matrix whose elements have zero mean and variance $N_0 = S / SNR$. The channel matrix $\mathbf{H}$ is complex Gaussian with zero mean and unit variance. The total transmitted power is scaled as $S$ in order to make the received Signal-to-Noise Ratio (SNR) $SNR$.

Let $\chi$ denote the signal set of $M$-QAM. Then the ML decoding of (\ref{eq:detected_matrix}) is obtained by
\begin{align}
\hat{\mathbf{X}} = \min_{s_j \in \chi} \| \mathbf{Y}- \mathbf{{\Lambda}X} \| ^2,
\end{align}
where $j \in \{1,\cdots,4\}$.

%% file: Analysis.tex
\section{Diversity Analysis} \label{sec:Analysis}

In this section, diversity analysis of GCMB is carried out by calculating the Pairwise Error Probability (PEP) between the transmitted codeword $\mathbf{X}$ and the detected codeword $\mathbf{\hat{X}}$. For ML decoding, the instantaneous PEP is represented as
\begin{align}
\mathrm{Pr} \left( \mathbf{X} \rightarrow \mathbf{\hat{X}} \mid \mathbf{H} \right) &= \mathrm{Pr} \left( \| \mathbf{Y} - \mathbf{\Lambda X} \| ^2 \geq \| \mathbf{Y} - \mathbf{\Lambda \hat{X}} \| ^2 \mid \mathbf{H} \right) \nonumber \\ &= \mathrm{Pr} \left( \xi \geq \| \mathbf{\Lambda} (\mathbf{X} - \mathbf{\hat{X}}) \| ^2 \mid \mathbf{H} \right), \label{eq:PEP}
\end{align}
where $\xi = \textrm{Tr} \{ - (\mathbf{X} - \mathbf{\hat{X}})^H \mathbf{\Lambda}^H \mathbf{N}- \mathbf{N}^H \mathbf{\Lambda} (\mathbf{X} - \mathbf{\hat{X}}) \}$. Since $\xi$ is a zero mean Gaussian random variable with variance $2 N_0 \| \mathbf{\Lambda}  (\mathbf{X} - \mathbf{\hat{X}}) \| ^2$, (\ref{eq:PEP}) is given by the $Q$ function as
\begin{align}
\mathrm{Pr} \left( \mathbf{X} \rightarrow \mathbf{\hat{X}} \mid \mathbf{H} \right) = Q \left( \sqrt{\frac{\| \mathbf{\Lambda} (\mathbf{X} - \mathbf{\hat{X}}) \| ^2}{2 N_0}}\right). \label{eq:PEP_2}
\end{align}
By using the upper bound on the $Q$ function $Q(x) \leq \frac{1}{2} e^{-x^2/2}$, the average PEP can be upper bounded as
\begin{align}
\mathrm{Pr} \left( \mathbf{X} \rightarrow \mathbf{\hat{X}} \right) &= E \left[ \mathrm{Pr} \left( \mathbf{X} \rightarrow
\mathbf{\hat{X}} \mid \mathbf{H} \right) \right] \nonumber \\ &\leq E \left[ \frac{1}{2} \exp \left(- \frac{\| \mathbf{\Lambda} (\mathbf{X} - \mathbf{\hat{X}}) \| ^2}{4 N_0}
\right) \right]. \label{eq:PEP_average}
\end{align}

Define $\mathbf{x}_1=[s_1, s_2]^T$ and $\mathbf{x}_2=[s_3, s_4]^T$. Then the Golden Code codeword can be represented as \cite{Belfiore_GC}
\begin{align}
\mathbf{X} = \textrm{diag}(\mathbf{Gx_1}) + \textrm{diag}(\mathbf{Gx_2})\mathbf{E}, \label{eq:golden_code_2}
\end{align}
where 
\begin{align*}
\mathbf{G}=\frac{1}{\sqrt{5}} \left[
\begin{array}{cc}
1+i\beta & {\alpha}-i \\
1+i\alpha & {\beta}-i
\end{array} \right],
\mathbf{E}= \left[
\begin{array}{cc}
0 & 1 \\
i & 0
\end{array} \right],
\end{align*}
and $\textrm{diag}[v_1, \cdots, v_S]$ denotes a diagonal matrix with diagonal entries $v_1, \cdots, v_S$. Let $\mathbf{g}_j$ with $j=1,\cdots, S$ denote the $j^{th}$ column of $\mathbf{G}^T$, then $\mathbf{g}^T_j$ denotes the $j^{th}$ row of $\mathbf{G}$. Equation (\ref{eq:golden_code}) is then rewritten as
\begin{align}
\mathbf{X} =  
\left[ \begin{array}{cc} 
\mathbf{g}^T_1\mathbf{x}_1 & \mathbf{g}^T_1\mathbf{x}_2 \\
i\mathbf{g}^T_2\mathbf{x}_2 & \mathbf{g}^T_2\mathbf{x}_1
\end{array} \right].
\label{eq:golden_code_3} 
\end{align}
Therefore,
\begin{align}
\mathbf{\Lambda X} =  
\left[ \begin{array}{cc} 
{\lambda}_1\mathbf{g}^T_1\mathbf{x}_1 & {\lambda}_1\mathbf{g}^T_1\mathbf{x}_2 \\
i{\lambda}_2\mathbf{g}^T_2\mathbf{x}_2 & {\lambda}_2\mathbf{g}^T_2\mathbf{x}_1
\end{array} \right].
\label{eq:Lambda_X} 
\end{align}
Then,
\ifCLASSOPTIONtwocolumn 
\begin{align}
\| \mathbf{\Lambda X} \|^2 &=  \textrm{Tr} \left\lbrace \mathbf{X}^H \mathbf{\Lambda}^H \mathbf{\Lambda X} \right\rbrace \nonumber \\ 
&= {\lambda}_1^2( | \mathbf{g}^T_1\mathbf{x}_1 |^2 + | \mathbf{g}^T_1\mathbf{x}_2 |^2) \nonumber \\
&+ {\lambda}_2^2( | \mathbf{g}^T_2\mathbf{x}_1 |^2 + | \mathbf{g}^T_2\mathbf{x}_2 |^2).
\label{eq:Lambda_X_square} 
\end{align}
\else
\begin{align}
\| \mathbf{\Lambda X} \|^2 &=  \textrm{Tr} \left\lbrace \mathbf{X}^H \mathbf{\Lambda}^H \mathbf{\Lambda X} \right\rbrace \nonumber \\ 
&= {\lambda}_1^2( | \mathbf{g}^T_1\mathbf{x}_1 |^2 + | \mathbf{g}^T_1\mathbf{x}_2 |^2) + {\lambda}_2^2( | \mathbf{g}^T_2\mathbf{x}_1 |^2 + | \mathbf{g}^T_2\mathbf{x}_2 |^2).
\label{eq:Lambda_X_square} 
\end{align}
\fi
Let $\hat{\mathbf{x}}_1=[\hat{s}_1, \hat{s}_2]$ and $\hat{\mathbf{x}}_2=[\hat{s}_3, \hat{s}_4]$ denote the detected symbol vectors. By replacing $\mathbf{x}_1$ and $\mathbf{x}_2$ in (\ref{eq:Lambda_X_square}) by $\mathbf{x}_1-\hat{\mathbf{x}}_1$ and $\mathbf{x}_2-\hat{\mathbf{x}}_2$, (\ref{eq:PEP_average}) is then rewritten as
\begin{align}
\mathrm{Pr} \left( \mathbf{X} \rightarrow \mathbf{\hat{X}} \right) \leq E \left[ \frac{1}{2} \exp \left(- \frac{{\rho}_1{\lambda}_1^2+{\rho}_2{\lambda}_2^2}{4 N_0} 
\right) \right], \label{eq:PEP_average_2}
\end{align}
where
\begin{align*}
\begin{split}
&{\rho}_1=( | \mathbf{g}^T_1(\mathbf{x}_1-\hat{\mathbf{x}}_1) |^2 + | \mathbf{g}^T_1(\mathbf{x}_2-\hat{\mathbf{x}}_2) |^2), \\
&{\rho}_2=( | \mathbf{g}^T_2(\mathbf{x}_1-\hat{\mathbf{x}}_1) |^2 + | \mathbf{g}^T_2(\mathbf{x}_2-\hat{\mathbf{x}}_2) |^2),
\end{split}
\end{align*}

The upper bound in (\ref{eq:PEP_average_2}) can be further bounded by employing a theorem from \cite{Park_UP_MPDF} which is specified below.

\begin{theorem}
Consider the largest $S \leq \min(N_t, N_r)$ eigenvalues $\mu_s$ of the uncorrelated central $N_r \times N_t$ Wishart matrix that are sorted in decreasing order, and a weight vector $\boldsymbol{\rho} = [\rho_1, \cdots, \rho_S]^T$ with
non-negative real elements. In the high SNR regime, an upper bound for the expression $E [ \exp (-\gamma
\sum_{s=1}^S \rho_s \mu_s ) ]$, which is used in the diversity analysis of a number of MIMO systems, is
\begin{align*}
E\left[ \exp \left( - \gamma \sum\limits_{s=1}^S \rho_s \mu_s
\right) \right] \leq \zeta \left( \rho_{min} \gamma
\right)^{-(N_r-\delta+1)(N_t-\delta+1)}
\end{align*}
where $\gamma$ is signal-to-noise ratio, $\zeta$ is a constant, $\rho_{min} = \min \{ \rho_1, \cdots, \rho_S \}$, and $\delta$
is the index to the first non-zero element in the weight vector.
\label{theorem:E_PEP}
\end{theorem}
\begin{IEEEproof}
See \cite{Park_UP_MPDF}.
\end{IEEEproof}

Note that ${\rho}_1 > 0$ and ${\rho}_2 > 0$, then $\delta=1$. By applying Theorem \ref{theorem:E_PEP} to (\ref{eq:PEP_average_2}), an upper bound of PEP is
\begin{align}
\mathrm{Pr} \left( \mathbf{X} \rightarrow \mathbf{\hat{X}} \right) &\leq \zeta \left( \frac{\min\{{\rho}_1,{\rho}_2\}}{4 N} SNR \right)^{-N_rN_t}.
\label{eq:PEP_PSB_final}
\end{align}
Hence, GCMB achieves the full diversity order of 4.

%% file: Decoding.tex
\section{Decoding} \label{sec:Decoding}

Equation (\ref{eq:Lambda_X}) shows that each element of $\mathbf{\Lambda X}$ is only related to $\mathbf{x}_1$ or $\mathbf{x}_2$. Consequently, the elements of $\mathbf{\Lambda X}$ can be divided into $S$ groups, where the $j^{th}$ group contains elements related to  $\mathbf{x}_j$ and $j=1, \cdots, S$. The input-output relation in (\ref{eq:detected_matrix}) then is decomposed into two equations as
\begin{align}
\begin{split}
&\mathbf{y}_1 = \left[
\begin{array}{c} 
Y_{1,1} \\
Y_{2,2}
\end{array} \right]  = \left[
\begin{array}{c}
{\lambda}_1\mathbf{g}^T_1\mathbf{x}_1 \\
{\lambda}_2\mathbf{g}^T_2\mathbf{x}_1
\end{array} \right] + \left[
\begin{array}{c} 
N_{1,1} \\
N_{2,2}
\end{array} \right], \\
&\mathbf{y}_2 = \left[
\begin{array}{c} 
Y_{1,2} \\
Y_{2,1}
\end{array} \right]  = \left[
\begin{array}{c}
{\lambda}_1\mathbf{g}^T_1\mathbf{x}_2 \\
i{\lambda}_2\mathbf{g}^T_2\mathbf{x}_2
\end{array} \right] + \left[
\begin{array}{c} 
N_{1,2} \\
N_{2,1}
\end{array} \right],
\end{split} \label{eq:deteced_symbol_decomposed}
\end{align} 
where $Y_{j,k}$ and $N_{j,k}$ denote the $(j,k)^{th}$ element of $\mathbf{Y}$ and $\mathbf{N}$ respectively. Let $\mathbf{n}_1=[N_{1,1}, N_{2,2}]^T$ and $\mathbf{n}_2=[N_{1,2}, N_{2,1}]^T$, then (\ref{eq:deteced_symbol_decomposed}) can be further rewritten as
\begin{align}
\begin{split}
&\mathbf{y}_1 = \mathbf{\Lambda G x_1} + \mathbf{n}_1, \\
&\mathbf{y}_2 = \mathbf{\Phi \Lambda G x_2} + \mathbf{n}_2,
\end{split} \label{eq:deteced_symbol_decomposed_2}
\end{align} 
where 
\begin{align*}
\mathbf{\Phi} = \left[\
\begin{array}{cc}
1 & 0 \\
0 & i
\end{array} \right].
\end{align*}

The input-output relation of (\ref{eq:deteced_symbol_decomposed_2}) implies that $\mathbf{x}_1$ and $\mathbf{x}_2$ can be decoded separately. Indeed, each relation of (\ref{eq:deteced_symbol_decomposed_2}) has a form similar to FPMB presented in \cite{Park_CPB}. In \cite{Azzam_SD_NLR}, \cite{Azzam_SD_RLR}, a reduced complexity SD is introduced. The technique takes advantage of a special real lattice representation, which introduces orthogonality between the real and imaginary parts of each symbol, thus enables employing rounding (or quantization) for the last two layers of the SD. When the dimension is $2\times2$, it achieves ML performance with the worst-case decoding complexity of $\mathcal{O}(M)$. This technique can be employed to decode $2\times2$ FPMB or GCMB. Moreover, lower decoding complexity can be achieved for GCMB because of the special property of the $\mathbf{G}$ matrix.

By using the QR decomposition of $\mathbf{\Lambda G}=\mathbf{QR}$, where $\mathbf{R}$ is an upper triangular matrix, and the matrix $\mathbf{Q}$ is unitary, (\ref{eq:deteced_symbol_decomposed_2}) is rewritten as 
\begin{align}
\begin{split}
&\tilde{\mathbf{y}}_1 = \mathbf{Q}^H \mathbf{y}_1 = \mathbf{R}\mathbf{x}_1 + \mathbf{Q}^H \mathbf{n}_1 = \mathbf{R}\mathbf{x}_1 + \tilde{\mathbf{n}}_1, \\
&\tilde{\mathbf{y}}_2 = \mathbf{Q}^H \mathbf{\Phi}^H \mathbf{y}_2 = \mathbf{R}\mathbf{x}_2 + \mathbf{Q}^H \mathbf{\Phi}^H \mathbf{n}_2 = \mathbf{R}\mathbf{x}_2 + \tilde{\mathbf{n}}_2.
\end{split} \label{eq:deteced_symbol_decomposed_3}
\end{align} 
Let $\mathbf{f}_j$ denote the $j^{th}$ column of 
\begin{align}
\mathbf{\Lambda G} =  
\left[ \begin{array}{cc} 
\lambda_1(1+i\beta) & \lambda_1({\alpha}-i) \\
\lambda_2(1+i\alpha) & \lambda_2({\beta}-i)
\end{array} \right],
\label{eq:Lambda_G} 
\end{align}
where $j={1,\cdots,S}$. The elements of $\mathbf{R}$ are calculated as
\begin{equation}
\begin{split}
&R_{1,1}=\| \mathbf{f}_1 \|, \\
&R_{1,2}=\frac{<\mathbf{f}_2,\mathbf{f}_1>}{\| \mathbf{f}_1 \|}=\frac{({\alpha}-{\beta})({\lambda}_1^2-{\lambda}_2^2)}{\| \mathbf{f}_1 \|}, \\
&R_{2,1}=0, \\
&R_{2,2}=\| \mathbf{d}_2 \|,
\label{eq:R_elements}
\end{split} 
\end{equation}
where $\mathbf{d}_2=\mathbf{f}_2-(\frac{ <\mathbf{f}_2,\mathbf{f}_1>} { \| \mathbf{f}_1 \| ^2 \| \mathbf{f}_2 \| ^2 }) \mathbf{f}_1$, $<\mathbf{f}_2,\mathbf{f}_1>=\mathbf{f}_2^H\mathbf{f}_1$, and $R_{j,k}$ denotes the $(j,k)^{th}$ element of $\mathbf{R}$. Based on (\ref{eq:R_elements}), the $\mathbf{R}$ matrix is proved to be real-valued, which means the real and imaginary parts of (\ref{eq:deteced_symbol_decomposed_3}) can be decoded separately. Consequently, (\ref{eq:deteced_symbol_decomposed_3}) can be decomposed further as 
\begin{align}
\begin{split}
&\Re \{\tilde{\mathbf{y}}_1\} = \mathbf{R}\Re \{\mathbf{x}_1\} + \Re \{\tilde{\mathbf{n}}_1\}, \\ 
&\Im \{\tilde{\mathbf{y}}_1\} = \mathbf{R}\Im \{\mathbf{x}_1\} + \Im \{\tilde{\mathbf{n}}_1\}, \\
&\Re \{\tilde{\mathbf{y}}_2\} = \mathbf{R}\Re \{\mathbf{x}_2\} + \Re \{\tilde{\mathbf{n}}_2\}, \\
&\Im \{\tilde{\mathbf{y}}_2\} = \mathbf{R}\Im \{\mathbf{x}_2\} + \Im \{\tilde{\mathbf{n}}_2\},
\end{split} \label{eq:deteced_symbol_decomposed_4}
\end{align} 
where $\Re \{ \mathbf{v}\}$ and $\Im \{ \mathbf{v}\}$ denote the real part of imaginary part of $\mathbf{v}$ respectively.

To decode each part of (\ref{eq:deteced_symbol_decomposed_4}), a two-level real-valued SD can be employed plus applying the rounding procedure for the last layer. As a result, the worst-case decoding complexity of GCMB is $\mathcal{O}(\sqrt{M})$. 

Previously, the ML decoding of GC was shown to have the worst-case complexity of $\mathcal{O}(M^{2.5})$ \cite{Sinnokrot_FMLD_GC}, \cite{Sinnokrot_GC_FD}. However, the above analysis proves that this complexity can be reduced substantially to only $\mathcal{O}(\sqrt{M})$ by applying GCMB when CSI is known at the transmitter. Furthermore, the complexity of GCMB is lower than the full-diversity full-multiplexing FPMB as well. The worst-case decoding complexity of $2\times2$ FPMB is $\mathcal{O}(M)$ with the decoding technique presented in \cite{Azzam_SD_NLR}, \cite{Azzam_SD_RLR}.

%% file: Extension.tex
\section{Extension to larger dimensions} \label{sec:Extension}
In \cite{Oggier_PSTBC}, the Golden Code is generalized to Perfect Space-Time Block Code (PSTBC) in dimensions $2$, $3$, $4$ and $6$, which have the full rate, the full diversity, nonvanishing minimum determinant for increasing spectral efficiency, uniform average transmitted energy per antenna, and good shaping of the constellation. In \cite{Elia_PSTBC}, PSTBCs have been generalized to any dimension. However, it is proved in \cite{Berhuy_PSTC} that particular perfect codes, yielding increased coding gain, only exist in dimensions $2$, $3$, $4$ and $6$. In this section, GCMB is generalized to larger dimensions of $3$, $4$, and $6$ by transmitting the corresponding PSTBC through multiple beamforming.

The codewords of a PSTBC are constructed as
\begin{equation}
\mathbf{X}=\sum_{j=1}^S{\textrm{diag}(\mathbf{G}\mathbf{x}_j)\mathbf{E}^{j-1}}
\label{eq:PSTBC}
\end{equation}
where $S$ is the system dimension, $\mathbf{G}$ is a $S\times S$ unitary matrix, $\mathbf{x}_j$ is a $S\times1$ vector whose elements are information symbols, and 
\begin{align*}
\mathbf{E} = \left[
\begin{array}{ccccc}
0 & 1 & \cdots & 0 & 0 \\
0 & 0 & 1 & \cdots & \vdots \\
\vdots & \vdots & \ddots & \ddots & \vdots \\
0 & \cdots & \cdots & \cdots & 1 \\
g & 0 & \cdots & 0 & 0
\end{array}
\right],
\end{align*}
with 
\begin{align*}
g = \left\lbrace 
\begin{array}{cc}
i, & S=2,4, \\
e^{\frac{2{\pi}i}{3}}, & S=3, \\
-e^{\frac{2{\pi}i}{3}}, & S=6.
\end{array} \right.
\end{align*}
The selection of the $\mathbf{G}$ matrix for different dimensions can be found in \cite{Oggier_PSTBC}.

In the sequel, the technique which transmits PSTBC through multiple beamforming is called Perfect Coded Multiple Beamforming (PCMB). Similarly to GCMB, the received signal of PCMB can be represented as (\ref{eq:detected_matrix}). PCMB achieves the full diversity order, which can be proved by analyzing the PEP in a similar way to Section \ref{sec:Analysis}.

Similarly to GCMB, the elements of $\mathbf{\Lambda X}$ for PCMB are related to only one of the $\mathbf{x}_j$, thus can be divided into $S$ groups. The received signal is then divided into $S$ parts, which can be represented as
\begin{align}
\mathbf{y}_j = \mathbf{\Phi}_j \mathbf{\Lambda G} \mathbf{x}_j + \mathbf{n}_j, 
\label{eq:deteced_symbol_decomposed_5}
\end{align} 
where $\mathbf{\Phi}_j=\textrm{diag}(\phi_{j,1}, \cdots, \phi_{j,S})$ is a diagonal unitary matrix whose elements satisfy
\begin{align*}
{\phi}_{j,k} = \left\lbrace 
\begin{array}{cc}
1, & 1 \leq k \leq S+1-j, \\
g, & S+2-j \leq k \leq S.
\end{array} \right.
\end{align*} 

By using the QR decomposition of $\mathbf{\Lambda G}=\mathbf{Q} \mathbf{R}$, where $\mathbf{R}$ is an upper triangular matrix, and the matrix $\mathbf{Q}$ is unitary, and moving $\mathbf{\Phi}_j\mathbf{Q}$ to the left hand, (\ref{eq:deteced_symbol_decomposed_5}) is rewritten as 
\begin{align}
\tilde{\mathbf{y}}_j = \mathbf{Q}^H \mathbf{\Phi}_j^H \mathbf{y}_j = \mathbf{R}\mathbf{x}_j + \mathbf{Q}^H \mathbf{\Phi}_j^H \mathbf{n}_j = \mathbf{R}\mathbf{x}_j + \tilde{\mathbf{n}}_j.
\label{eq:deteced_symbol_decomposed_6}
\end{align}

For the dimension of $S=4$, the $\mathbf{R}$ matrix in (\ref{eq:deteced_symbol_decomposed_6}) is real-valued,  which can be proved in a similar way to Section \ref{sec:Decoding}. Consequently, the real part and the imaginary part of $\mathbf{x}_j$ can be decoded separately, in a similar way to GCMB. Real-valued SD with the last layer rounded can be employed to decode (\ref{eq:deteced_symbol_decomposed_6}). The worst-case decoding complexity of PCMB is then $\mathcal{O}(M^{1.5})$. Regarding MIMO systems using PSTBC, the worst-case decoding complexity is $\mathcal{O}(M^{13.5})$ by using a similar decoding technique to \cite{Sinnokrot_FMLD_GC}, \cite{Sinnokrot_GC_FD}. For FPMB, ML decoding can be achieved by using SD based on the real lattice representation in \cite{Azzam_SD_NLR}, \cite{Azzam_SD_RLR}, plus quantization of the last two layers, and the worst-case complexity is $\mathcal{O}(M^{3})$.

For the $S=3$  dimension case, the $\mathbf{R}$ matrix is complex-valued. Therefore, the real and the imaginary parts of $\mathbf{x}_j$ cannot be decoded separately, unlike the case of $S=2,4$. Moreover, since the M-HEX constellation \cite{Forney_EM} is used instead of M-QAM, a complex-valued SD is needed. The worst-case decoding complexity of PCMB is then $\mathcal{O}(M^{3})$. In the case of general MIMO systems using PSTBC, the worst-case decoding complexity is $\mathcal{O}(M^{9})$. For FPMB, the worst-case decoding complexity is $\mathcal{O}(M^{2})$.

In the case of $S=6$, which is similar to $S=3$, the $\mathbf{R}$ matrix is complex-valued, and M-HEX signals are transmitted. Consequently, the worst-case decoding complexity is $\mathcal{O}(M^{6})$, while the worst-case decoding complexity of general MIMO systems using PSTBC and FPMB are $\mathcal{O}(M^{36})$ and $\mathcal{O}(M^{5})$ respectively.

%% file: Results.tex
\section{Simulation Results} \label{sec:Results}

Considering $2\times2$ systems, Fig. \ref{fig:ber_snr_2x2} shows BER-SNR performance comparison of GCMB, FPMB, and general MIMO systems using the Golden Code, which is denoted by GC, for different modulation schemes. The constellation precoder for FPMB is selected as the best one introduced in \cite{Park_CPB}. Simulation results show that GCMB, GC, and FPMB, with the worst-case decoding complexity of $\mathcal{O}(\sqrt{M})$, $\mathcal{O}(M^{2.5})$, and $\mathcal{O}(M)$, respectively, achieve very close performance for all of $4$-QAM, $16$-QAM, and $64$-QAM. The performance differences among these three are less than $1$dB, and become smaller when the modulation alphabet size increases. 

In the case of $4\times4$ systems, Fig. \ref{fig:ber_snr_4x4} shows BER-SNR performance comparison of PCMB, FPMB, and general MIMO systems using the PSTBC, which is denoted by PC, for $4$-QAM and $16$-QAM. The constellation precoder for FPMB is also chosen as the best one in \cite{Park_CPB}. Simulation results show that PCMB has approximately $3$dB and $1$dB performance degradations compared to PC and FPMB, respectively, and the degradations decrease as the modulation alphabet size increases. However, the performance compromises of PCMB trade off with reductions of the worst-case decoding complexity for PC and FPMB from $\mathcal{O}(M^{13.5})$ and $\mathcal{O}(M^{3})$ to only $\mathcal{O}(M^{1.5})$, respectively. 

\ifCLASSOPTIONonecolumn
\begin{figure}[!m]
\centering
\scalebox{.7}{\includegraphics{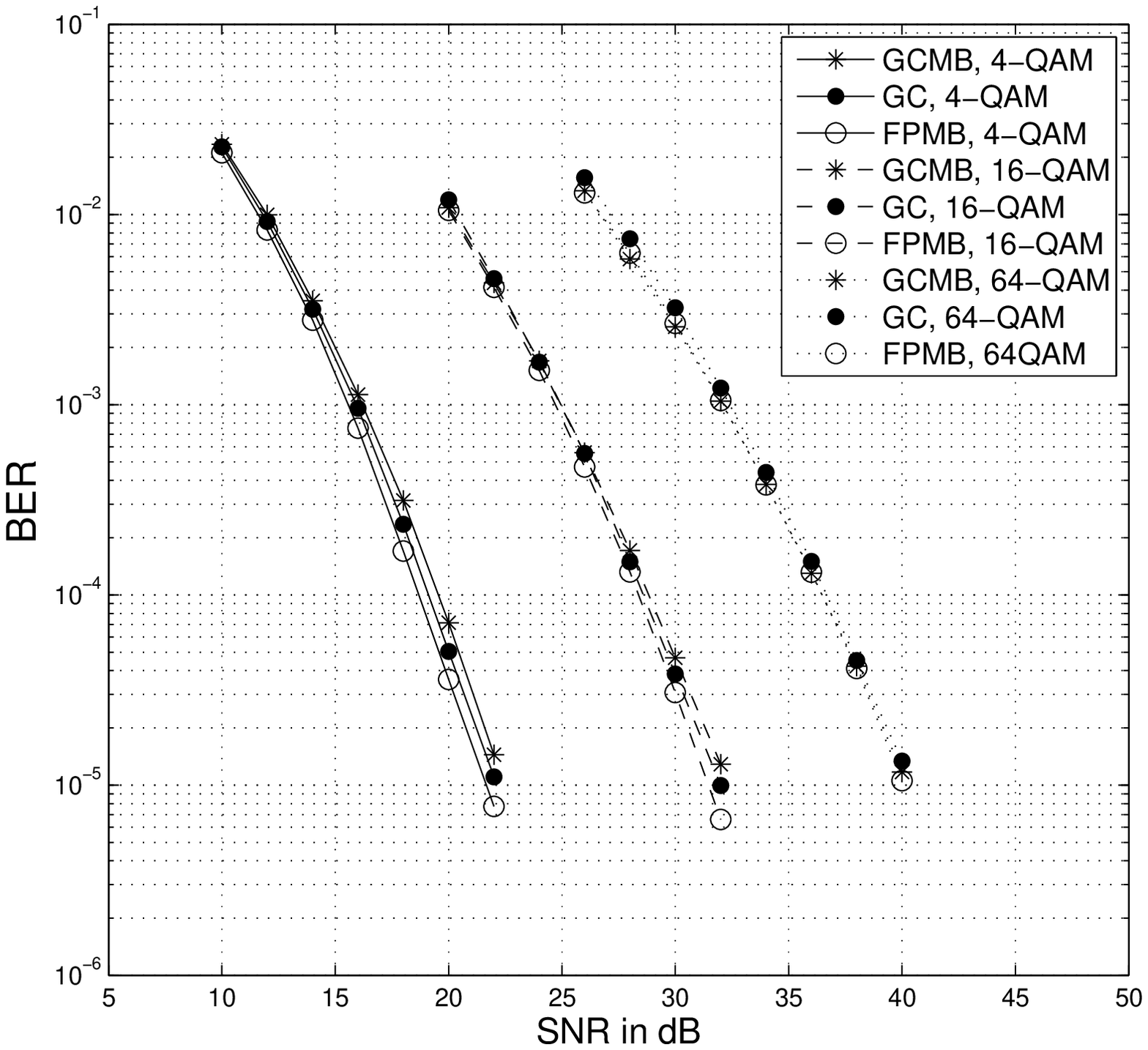}}
\caption{BER vs. SNR for GCMB, GC and FPMB for $2\times2$ systems.}
\label{fig:ber_snr_2x2}
\end{figure}

\begin{figure}[!m]
\centering
\scalebox{.7}{\includegraphics{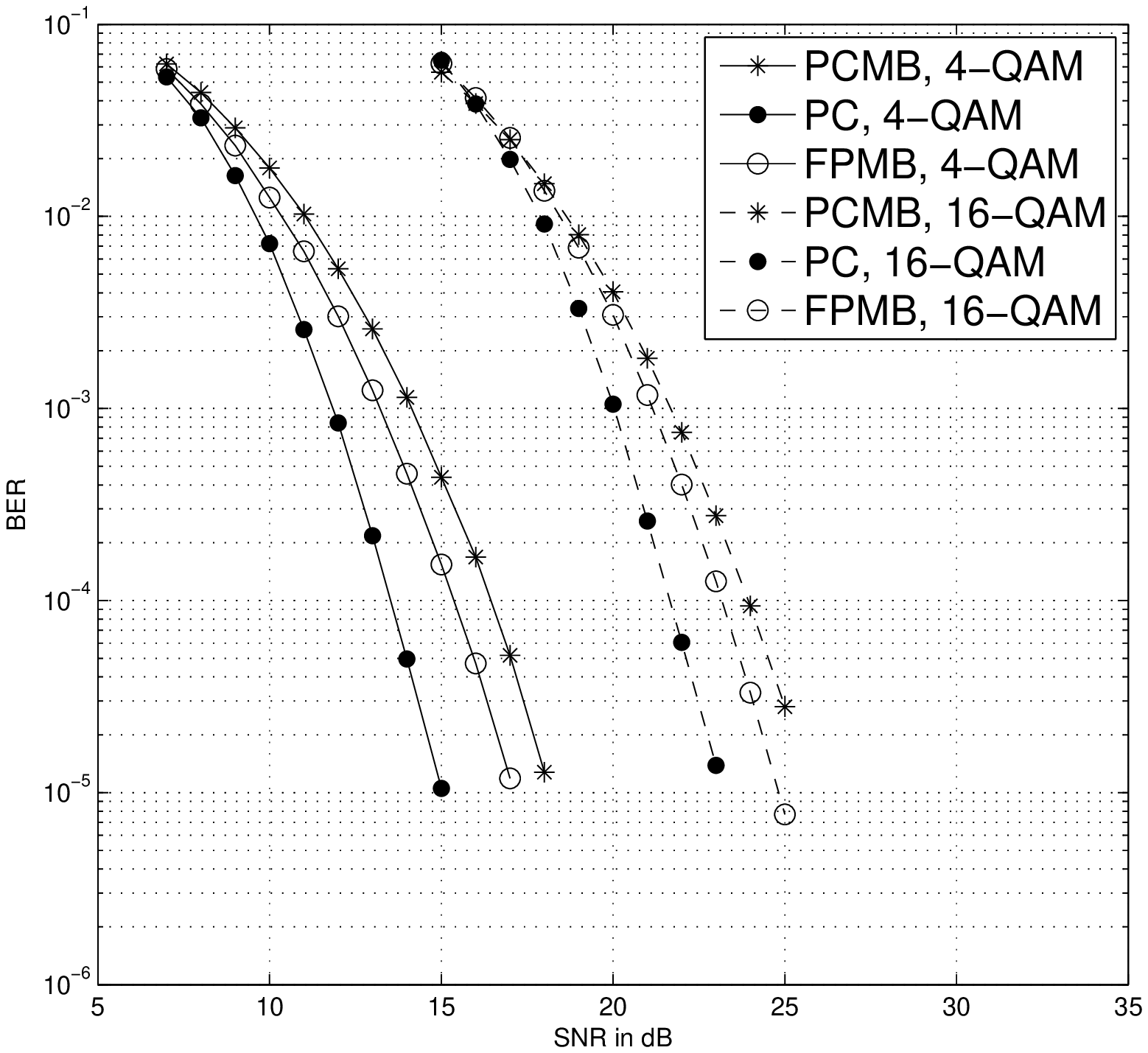}}
\caption{BER vs. SNR for PCMB, PC and FPMB for $4\times4$ systems.}
\label{fig:ber_snr_4x4}
\end{figure}

\else
\begin{figure}[!t]
\centering
\scalebox{.5}{\includegraphics{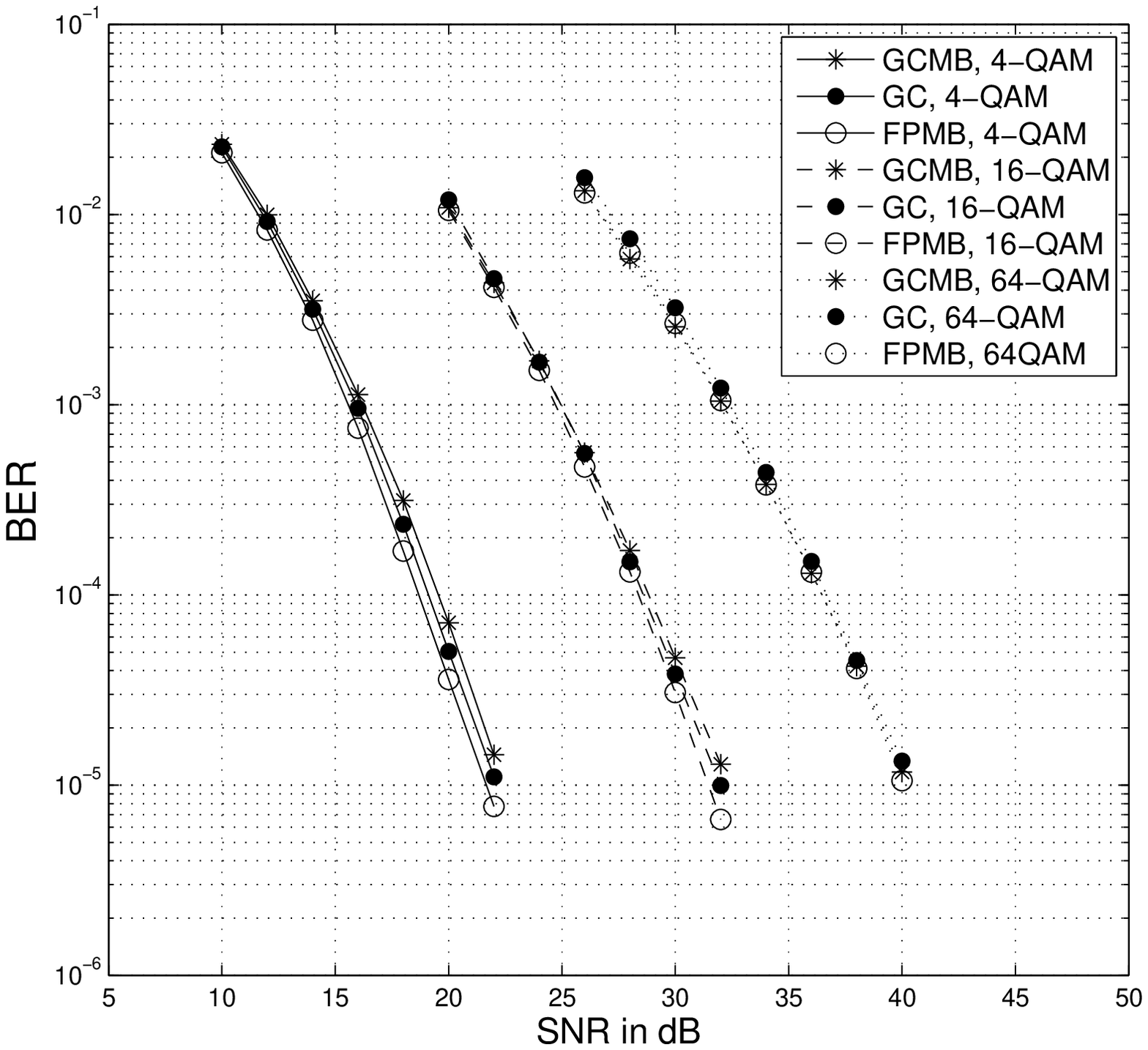}}
\caption{BER vs. SNR for GCMB, GC and FPMB for $2\times2$ systems.}
\label{fig:ber_snr_2x2}
\end{figure}

\begin{figure}[!t]
\centering
\scalebox{.5}{\includegraphics{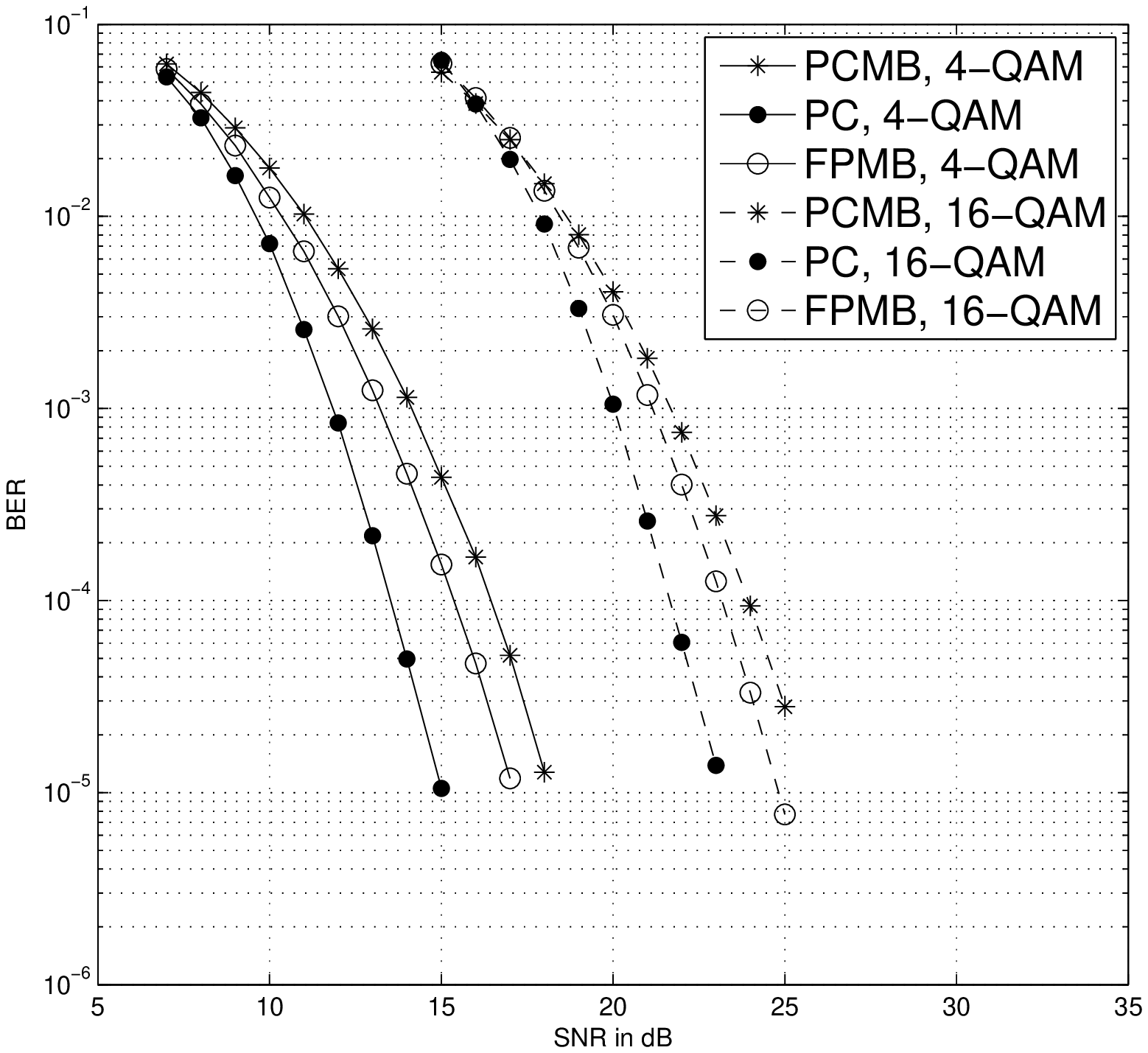}}
\caption{BER vs. SNR for PCMB, PC and FPMB $4\times4$ systems.}
\label{fig:ber_snr_4x4}
\end{figure}

\fi

%% file: Conclusion.tex
\section{Conclusion} \label{sec:Conclusion}

In this paper, GCMB which combines the Golden Code and multiple beamforming technique is proposed. It is shown that GCMB achieves full-diversity, full-rate, and low decoding complexity. Compared to general MIMO systems using the Golden Code, GCMB has similar performance while the worst-case decoding complexity is reduced from $\mathcal{O}(M^{2.5})$ to only $\mathcal{O}(\sqrt{M})$, when square $M$-QAM is used. The substantial complexity reduction benefits from the knowledge of CSI at the transmitter. Moreover, the complexity of GCMB is also lower than $2\times2$ FPMB, which is a full-diversity full-rate beamforming technique without channel coding, with the worst-case decoding complexity of $\mathcal{O}(M)$. Similarly, GCMB and FPMB have very close performance.

GCMB is generalized to PCMB in dimensions $3$, $4$, and $6$. PCMB combines PSTBC with multiple beamforming. Similarly to GCMB, PCMB reduces the worst-case decoding complexity of general MIMO systems using PSTBC from $\mathcal{O}(M^{9})$, $\mathcal{O}(M^{13.5})$, and $\mathcal{O}(M^{36})$ to $\mathcal{O}(M^{3})$, $\mathcal{O}(M^{1.5})$, and $\mathcal{O}(M^{6})$ in dimensions $3$, $4$, and $6$, respectively. Compared to FPMB, the worst-case decoding complexity of PCMB is lower than $\mathcal{O}(M^{3})$ of FPMB in dimension $4$, while due to the complex-valued $\mathbf{R}$ matrix and the HEX signals, higher than $\mathcal{O}(M^{2})$ and $\mathcal{O}(M^{5})$ of FPMB in dimensions $3$ and $6$, respectively.